# A Generalized Empirical Likelihood Approach for Two-Group Comparisons Given a U-Statistic Constraint

Jihnhee Yu*, Luge Yang, Albert Vexler, Alan D. Hutson

Department of Biostatistics, University at Buffalo, State University of New York.

*Corresponding author, jinheeyu@buffalo.edu

**Abstract**

We investigate a generalized empirical likelihood approach in a two-group setting where the constraints on parameters have a form of U-statistics. In this situation, the summands that consist of the constraints for the empirical likelihood are not independent, and a weight of each summand may not have a direct interpretation as a probability point mass, dissimilar to the common empirical likelihood constraints based on independent summands. We show that the resulting empirical likelihood ratio statistic has a weighted $\chi^2$ distribution in the univariate case and a combination of weighted $\chi^2$ distributions in the multivariate case. Through an extensive Monte-Carlo study, we show that the proposed methods applied for some well-known U-statistics have robust Type I error control under various underlying distributions including cases with a violation of exchangeability under null hypotheses. For the application, we employ the proposed methods to test hypotheses in crossover designs demonstrating an adaptability of the proposed methods in various hypothesis tests.

## 1. Introduction

In this paper we investigate the asymptotic properties of the empirical likelihood (EL) ratio test in a two-group setting, where the constraints have the form of general U-statistics. In this situation, the summands (U-statistic kernel) that consist of the constraints for the empirical likelihood are not independent, thus the common EL approach based on independent summands may not be applicable. In this investigation, we do not manipulate the summands to be pseudo-independent (e.g., [1]) nor do we artificially force this into a one-sample problem (e.g., [2]). The proposed approach is easy to implement for various U-statistics type of constraints. We demonstrate the applicability of the proposed methods by applying it to some well-known U-statistic-based test statistics and show their workability and accuracy.

Since being introduced into the statistical literature, the EL approach [3] has demonstrated its practical applicability via extensions to a variety of statistical problems (e.g., [4-6]). The EL approach is a data-driven likelihood function, which is able to incorporate known constraints on parameters in an inferential setting under both the null and alternative hypotheses. EL hypothesis tests maintain a pre-specified Type I error rate relatively well even under violations of the exchangeability assumptions (e.g., [7]). In comparison with classical testing methods based on normal approximations, the EL test statistic does not rely on symmetric rejection regions, thus giving rise to more accurate tests [8].

Typically, the EL construction is based on independent observations. Specifically, with a simple expectation constraint in a one-sample setting based on $X_i$ $(i=1,\cdots,n)$, the EL function $\prod_{i=1}^{n} p_i$ is constructed as

$$\max\left\{\prod_{i=1}^{n} p_i : \sum_{i=1}^{n} p_i = 1, \sum_{i=1}^{n} p_i \phi(x_i, \theta) = 0\right\}, \qquad (1)$$

where $p_i$ ($0 \leq p_i \leq 1$) represents a probability point mass corresponding to an independent data point, $\phi(x_i, \theta)$ is a relevant summand as a function of observations corresponding to a constraint of interest and $\theta$ is a parameter of interest (e.g., parameter under the null hypothesis). For an application to U-statistic type constraints, $\phi(\cdot)$ in (1) can be replaced by a relevant U-statistic kernel and the summation can be carried out over all permutations of the data. While the EL approach incorporates constraints composed of independent summands, the kernels of a U-statistic corresponding to the collection of the permutations of the indices are not independent. Also, $p_i$ or weight of a summand in (1) may not be directly interpreted as a probability point mass corresponding to each observation as we will explain in Section 2.

Recent investigations that have dealt with applying the EL approach to U-statistics are found primarily in a one-sample setting. Wood et al. [9] considered the EL approach to U-statistics via a bootstrap calibration. Most recently, a comprehensive investigation of the EL method involving U-statistics in a one-sample setting was carried out by Yuan et al. [10].

In a two-sample setting, EL approaches based on general U-statistic constraints have not been investigated except in the article by Jing et al. [1]. Some articles introduce an application of classical EL methods to the analysis of the receiver operating characteristic curves [2, 11], where the kernel of a U-statistic was manipulated to use the classical EL methods. These methodologies do not provide an EL approach applied to general U-statistic constraints. For example, Claeskens et al. [11] considered the receiver operating characteristic (ROC) curve problem, where constraints were set up for sensitivity and specificity. In order to investigate the area under the ROC curve, the whole range of specificities (between 0 and 1) needed to be examined in a point-wise manner [11]. Qin and Zhou [2], in their investigation of the test of the area under an ROC curve, constructed a constraint for one group so that the test could be performed in a similar manner to the one-sample EL method via sample estimates of specificity and the "plug-in" method (e.g., [12]). For an application to more general U-statistics, Jing et al. [1] obtained an empirical likelihood based on jackknife pseudo-values that were conjectured to be independent. In this paper we obtain the EL test statistic with general two-sample U-statistics, where no such forced modifications described above are used. Thus, our approach and results are simple in that they do not require much alteration to implement for various U-statistics.

This paper has the following structure. In Section 2 we develop the EL tests for two-groups based on general U-statistic constraints. In Section 3 we consider adaptations of the proposed methods to tests based on some popular U-statistics. In Section 4 we investigate the performance of the proposed methods developed in Section 3 through an extensive Monte-Carlo study under various underlying distributions and sample sizes. In Section 5 we apply the proposed methods to test various hypotheses resulting from crossover designs. Finally, Section 6 is devoted to discussions and concluding remarks.

## 2. Main Results

Let $X_1, \cdots, X_{n_1}$ be $n_1$ independent observations from group 1 with unknown distribution function $F(x) = p(X_i < x)$, and let $Y_1, \cdots, Y_{n_2}$ be $n_2$ independent observations from group 2 with unknown distribution function $G(y) = p(Y_j < y)$. Note that, for simplicity, we start by dealing with the univariate random variables primarily in the two-group comparison setting. However, the approach based on the univariate variables is easily generalized to multivariate random variables, which we will develop later in this section. Let $m_1 \leq n_1$ and $m_2 \leq n_2$, and $h(X_1, \cdots, X_{m_1}, Y_1, \cdots, Y_{m_2}) = h(\mathbf{X}, \mathbf{Y})$ be a symmetric kernel of a U-statistic for the two-groups. Also, let $E_{F,G}(h(\mathbf{X}, \mathbf{Y})) = \theta$, and $E_{F,G}(h(\mathbf{X}, \mathbf{Y}) - \theta)^2 = \eta^2$, where $E_{F,G}$ denotes the expectation with respect to the true underlying distributions $F$ and $G$. The corresponding U-statistic estimator of $\theta$ is given as

$$\hat{\theta} = \frac{1}{\binom{n_1}{m_1}\binom{n_2}{m_2}} \sum_{i,j \in S} h(\mathbf{X}_i, \mathbf{Y}_j), \tag{2}$$

where $i = (i_1, \ldots, i_{m_1})$, $j = (j_1, \ldots, j_{m_2})$, and $S = \{i, j: 1 \le i_1 < \ldots < i_{m_1} \le n_1,\ 1 \le j_1 < \ldots < j_{m_2} \le n_2\}$ denotes all permutations of the $m_k$ indices for group $k$ ($k = 1, 2$). The U-statistic in (2) is known to be an unbiased estimator of $\theta$. Note that, given all $h(\mathbf{X}_i, \mathbf{Y}_j) = 1$, we have

$$\left[\binom{n_1}{m_1}\binom{n_2}{m_2}\right]^{-1} \sum_{i,j \in S} 1 = 1. \tag{3}$$

Following a conventional notation [13], let us define

$$h_{c_1, c_2}(X_1, \ldots, X_{c_1}, Y_1, \ldots, Y_{c_2}) = \int \ldots \int h(u_1, \ldots, u_{m_1}, v_1, \ldots, v_{m_2}) \prod_{i=1}^{c_1} (d\delta_{x_i}(u_i) - dF(u_i))$$
$$\times \prod_{i=c_1+1}^{m_1} dF(u_i) \prod_{j=1}^{c_2} (d\delta_{y_j}(v_j) - dG(v_j)) \prod_{j=c_2+1}^{m_2} dG(v_j), \tag{4}$$

for some integers $c_1 (\le m_1)$ and $c_2 (\le m_2)$, where $\delta_x$ denotes the distribution function of a single point mass at $x$. In (4), if $c_i = 0$, the product involving $\delta_x$ is omitted. The asymptotic variance of $(n_1 + n_2)^{1/2} \hat{\theta}$ [13] has the form of

$$\frac{m_1^2(n_1+n_2)}{n_1} \sigma_{1,0}^2 + \frac{m_2^2(n_1+n_2)}{n_2} \sigma_{0,1}^2, \tag{5}$$

where $\sigma_{c_1,c_2}^2 = Var(h_{c_1,c_2}(X_1, \ldots, X_{c_1}, Y_1, \ldots, Y_{c_2}))$.

Now, consider the null hypothesis of interest:

$$H_0: E_{F,G}(h(\mathbf{X}, \mathbf{Y})) = \theta_0. \tag{6}$$

For the empirical constraint corresponding to (6), we use a weight $w_{ij}$ instead of $\left[\binom{n_1}{m_1}\binom{n_2}{m_2}\right]^{-1}$ in (2) as

$$\sum_{i,j \in S} w_{ij} h(\mathbf{X}_i, \mathbf{Y}_j) = \theta_0, \qquad (7)$$

where $\sum_{i,j \in S} w_{ij} = 1$, and $w_{ij} \geq 0$, analogous to what is presented in the equation (3). The value of $w_{ij}$ will be obtained to maximize the EL function with a constraint imposed on parameters of interest such as (6). We note that, without the constraint, $w_{ij} = \left[\binom{n_1}{m_1}\binom{n_2}{m_2}\right]^{-1}$. If $m_1 = m_2 = 1$, $w_{ij}$ has a direct interpretation as a probability point mass for $(x_i, y_j)$, that is, $w_{ij}$ is the estimated probability mass $\hat{P}(X = x_i, Y = y_j)$ similar to the typical interpretation of the weights in the EL approach [14] such that $\sum_{i=1}^{n_1} \hat{P}(X = x_i) = \sum_{j=1}^{n_2} \hat{P}(Y = y_j) = 1$. For the other values of $m_1$ and $m_2$, the correspondence of $w_{ij}$ to a probability point mass is generally not true. For example, consider a symmetric kernel with $m_1 = 1$ and $m_2 = 2$. Now, let each $w_{ij}$ correspond to the empirical estimate of the probability $P(\mathbf{X}_i, \mathbf{Y}_j)$, i.e., $\hat{P}(X_i, Y_{j_1}, Y_{j_2})$. Then,

$\sum_{i,j \in S} w_{ij} = \sum_{i=1}^{n_1} \hat{P}(X_i) \sum_{j_1=1}^{n_2-1} \sum_{j_2 > j_1}^{n_2} \hat{P}(Y_{j_1}, Y_{j_2})$ which is smaller than

$\sum_{i=1}^{n_1} \hat{P}(X_i) \sum_{j_1=1}^{n_2} \sum_{j_2=1}^{n_2} \hat{P}(Y_{j_1}, Y_{j_2}) = 1$. Thus, since we let $\sum_{i,j \in S} w_{ij} = 1$, each $w_{ij}$ does not correspond to a probability point mass.

The EL function under $H_0$ is defined as

$$L_{\theta_0} = \max_{w_{ij}} \prod_{i,j \in S} w_{ij}, \qquad (8)$$

which is subject to $\sum_{i,j \in S} w_{ij} = 1$, $0 \leq w_{ij} \leq 1$ and the constraint (7). The EL function under the alternative hypothesis $H_1$ (i.e., not $H_0$) is obtained similarly in an unconstrained manner per the constraint (7). It is easily shown that the EL function under $H_1$ is

$$L_{\theta_1} = \left[\binom{n_1}{m_1}\binom{n_2}{m_2}\right]^{-\binom{n_1}{m_1}\binom{n_2}{m_2}}. \qquad (9)$$

Obtaining (8) is carried out by maximizing $\sum_{i,j \in S} \log w_{ij}$ using the method of Lagrange multipliers with conditions $\sum_{i,j \in S} w_{ij} = 1$ and the constraint (7). This step gives us

$$w_{ij} = \left\{\binom{n_1}{m_1}\binom{n_2}{m_2}(1 + \lambda^*(h(\mathbf{X}_i, \mathbf{Y}_j) - \theta_0))\right\}^{-1}, \; i, j \in S, \qquad (10)$$

where $\lambda^*$ satisfies

$$\sum_{i,j \in S} \frac{h(\mathbf{X}_i, \mathbf{Y}_j) - \theta_0}{1 + \lambda^*(h(\mathbf{X}_i, \mathbf{Y}_j) - \theta_0)} = 0. \qquad (11)$$

By (8), (9) and (10), we obtain the EL ratio type test as

$$R(\theta_0) = \prod_{i,j \in S}\left\{1 + \lambda^*(h(\mathbf{X}_i, \mathbf{Y}_j) - \theta_0)\right\}^{-1},$$

and the corresponding EL log-likelihood ratio test is

$$l(\theta_0) = -2 \log R(\theta_0) = 2 \sum_{i,j \in S} \log(1 + \lambda^*(h(\mathbf{X}_i, \mathbf{Y}_j) - \theta_0)). \qquad (12)$$

Assume that $E_{F,G}|h(\mathbf{X}, \mathbf{Y})|^\alpha$ exists up to some positive $\alpha$ and $\theta_0$ is inside the convex hull of points given by the set of points $h(\mathbf{X}_i, \mathbf{Y}_j)$, $i, j \in S$. Also, assume that $n = (n_1 + n_2) \to \infty$ and $\frac{n_1}{n_1+n_2} \to r$ for asymptotic results. Let $V(\hat{\theta}) = \sigma^2$ denote the variance of $\hat{\theta}$. Then, we can show the following result:

*Theorem 1:* Under $H_0$, $\gamma(\theta_0)l(\theta_0)$ converges in distribution to $\chi_1^2$ distribution, where

$$\gamma(\theta_0) = \eta^2 \bigg/ \left[\binom{n_1}{m_1}\binom{n_2}{m_2}\sigma^2\right].$$

The proof of Theorem 1 is given in the Appendix.

To compute $w_{ij}$ in (10), we need to solve the nonlinear equation given in (11), which can be carried out using commonly available optimization packages. An initial value can be given by equating the first order expansion of the left-sided equation of (11) to 0 (see the proof of Theorem 1 in the Appendix for details). The appropriate solution is given within some neighborhood of the initial value.

The extension of the EL ratio test to multivariate random variables follows the univariate development outlined above. Now, consider $X_1, \cdots, X_{n_1}$ and $Y_1, \cdots, Y_{n_2}$ as the $p$-variate random vectors. We have that $E_{F,G}((h(\mathbf{X},\mathbf{Y}) - \theta)(h(\mathbf{X},\mathbf{Y}) - \theta)^T) = H$ and $\Sigma$ is the covariance matrix of $\hat{\theta}$. Note that the null hypothesis is given as $H_0 : E_{F,G}(h(\mathbf{X},\mathbf{Y})) = \theta_0$, where $\theta_0$ is a $q \times 1$ vector ($q \leq p$). The corresponding log-likelihood ratio test has the form of

$$l(\theta_0) = 2 \sum_{i,j \in S} \log(1 + \lambda^{*T}(h(\mathbf{X}_i, \mathbf{Y}_j) - \theta_0)), \tag{13}$$

where $q \times 1$ vector $\lambda^*$ satisfies

$$\sum_{i,j \in S} \frac{h(\mathbf{X}_i, \mathbf{Y}_j) - \theta_0}{1 + \lambda^{*T}(h(\mathbf{X}_i, \mathbf{Y}_j) - \theta_0)} = 0, \tag{14}$$

where 0 is the $q \times 1$ zero vector. As in the univariate setting, assume that $E_{F,G} \|h(\mathbf{X},\mathbf{Y})\|^\alpha$ exists, and $\theta_0$ is inside the convex hull given by the set of points $h(\mathbf{X}_i, \mathbf{Y}_j)$, $i, j \in S$. Then, we have the following result:

*Theorem 2: Under $H_0$, $l(\theta_0)\left[\binom{n_1}{m_1}\binom{n_2}{m_2}\right]^{-1}$ in (13) converges in distribution to $\sum_{k=1}^{q} c_k \chi_{1k}^2$ where $c_k$'s are the eigenvalues of $H^{-1}\Sigma$ and $\chi_{1k}^2$'s are independent $\chi_1^2$ random variables.*

The proof of Theorem 2 is similar to that of Theorem 1. A brief sketch of the proof is given in the Appendix.

In the following section, we will illustrate a few adaptations of the developed theoretical results on some popular two-group U statistics.

## 3. Adaptations to Various Tests

Applying the proposed method requires a relevant kernel of a U-statistic and variance estimates of the U-statistic. In general, the variance estimates of U-statistics can be obtained through resampling techniques [13]. If the analytical form of the variance estimate can be obtained, a straightforward application of the proposed method is possible as demonstrated in the following examples. Some relevant simulation results regarding the methods developed in this section are given in Section 4.

### 3.1 ROC curve analysis

The ROC curve has been used as an important tool to examine the discriminant ability of a biomarker for separating individuals with a certain disease from those without the disease. The ROC curve analysis looks at the probability $P(Y > X)$ where $X$ and $Y$ are random variables representing two different populations. Suppose $X_1, \cdots, X_{n_1}$ are $n_1$ independent observations from population 1 and $Y_1, \cdots, Y_{n_2}$ are $n_2$ independent observations from population 2. Also, let $X_i \sim X$ and $Y_i \sim Y$. The corresponding U-statistic for estimating the probability $\varsigma = P(Y > X)$ is given by

$$\hat{\varsigma}_{X,Y} = \frac{1}{n_1 n_2} \sum_{i=1}^{n_1} \sum_{j=1}^{n_2} \phi(X_i, Y_j), \quad (15)$$

where $\phi(X_i, Y_j) = I(X_i < Y_j)$ and $I$ denotes the indicator function. The statistic $\hat{\varsigma}_{X,Y}$ is essentially the Wilcoxon-Mann-Whitney U-statistic, which estimates the area under the estimated ROC curve (referred to as AUC). Let $\varsigma_0$ denote the AUC under $H_0$. We construct the EL ratio test as

$$R(\varsigma_0) = \frac{\prod_{i=1}^{n_1} \prod_{j=1}^{n_2} w_{ij}}{(n_1 n_2)^{-n_1 n_2}}, \quad (16)$$

where $w_{ij}$'s satisfy

$$Sup\left\{ \prod_{i=1}^{n_1} \prod_{j=1}^{n_2} w_{ij} : \sum_{i=1}^{n_1} \sum_{j=1}^{n_2} w_{ij} = 1, \sum_{i=1}^{n_1} \sum_{j=1}^{n_2} w_{ij}(\phi_{ij} - \varsigma_0) = 0 \right\},$$

and $\phi_{ij} = \phi(X_i, Y_j)$. Sen [15] provided a consistent estimate of the variance for the U-statistic $\hat{\varsigma}$, which we can incorporate in our development. Let

$$S_{10}^2 = \frac{1}{n_1 - 1} \sum_{i=1}^{n_1} \left(V_{10}(X_i) - \hat{\varsigma}_{X,Y}\right)^2, \text{ and } S_{01}^2 = \frac{1}{n_2 - 1} \sum_{j=1}^{n_2} \left(V_{01}(Y_j) - \hat{\varsigma}_{X,Y}\right)^2, \quad (17)$$

where $V_{10}(X_i) = \frac{1}{n_2} \sum_{j=1}^{n_2} \phi(X_i, Y_j)$ for $i = 1, \cdots, n_1$ and $V_{01}(Y_j) = \frac{1}{n_1} \sum_{i=1}^{n_1} \phi(X_i, Y_j)$ for $j = 1, \cdots, n_2$.

Then, the variance estimate of $\hat{\varsigma}_{X,Y}$ [15] is given as

$$\hat{V}(\hat{\varsigma}_{X,Y}) = \frac{n_1 + n_2}{n_1 n_2} \left( \frac{n_2 S_{10}^2 + n_1 S_{01}^2}{n_1 + n_2} \right). \quad (18)$$

Based on Theorem 1, equation (18) and the consistent estimate of $V(\hat{\varsigma}_{X,Y})$, we have the following result:

*Corollary 1:* Under $H_0: \varsigma = \varsigma_0$, $-2\log R(\varsigma_0) \dfrac{\sum_{i=1}^{n_1}\sum_{j=1}^{n_2}(\phi_{ij}-\varsigma_0)^2}{(n_1 n_2)^2 \hat{V}(\hat{\varsigma}_{X,Y})}$ converges in distribution to $\chi_1^2$ distribution.

The value $\hat{\varsigma}_{X,Y}$ in (17) can be replaced by $\varsigma_0$ under $H_0$. We demonstrate this approach using simulations as shown in Section 4.

3.2 Generalization for comparing two correlated AUC's

Suppose that individuals from population 1 provide independent observations $X_i = (X_{1i},...,X_{pi})^T$, $i=1,...,n_1$ and individuals from population 2 provide independent observations $Y_j = (Y_{1j},...,Y_{pj})^T$, $j=1,...,n_2$. Then we may be interested in testing generalized differences in AUC's given by the hypothesis,

$$H_0: p(\ell_1^T X < \ell_2^T Y) - p(\ell_3^T X < \ell_4^T Y) = \delta_0, \tag{19}$$

where $X \sim X_i$ and $Y \sim Y_i$, and $\ell_i$'s are some contrast vectors. The corresponding U-statistic $\hat{\delta}$ is

$$\hat{\delta} = \frac{1}{n_1 n_2}\sum_{i=1}^{n_1}\sum_{j=1}^{n_2}\{\phi(l_1^T X_i, l_2^T Y_j) - \phi(l_3^T X_i, l_4^T Y_j)\},$$

under the notation in Section 3.1. We construct the EL ratio test as

$$R(\delta_0) = \frac{\prod_{i=1}^{n_1}\prod_{j=1}^{n_2} w_{ij}}{(n_1 n_2)^{-n_1 n_2}}, \tag{20}$$

where $w_{ij}$'s satisfy

$$Sup\left\{\prod_{i=1}^{n_1}\prod_{j=1}^{n_2} w_{ij} : \sum_{i=1}^{n_1}\sum_{j=1}^{n_2} w_{ij} = 1, \sum_{i=1}^{n_1}\sum_{j=1}^{n_2} w_{ij}(\phi_{12_{ij}} - \phi_{34_{ij}} - \delta_0) = 0\right\},$$

and $\phi_{kk'_{ij}} = I(\ell_k^T X_i < \ell_{k'}^T Y_j)$. Since $\ell_i^T X$ and $\ell_j^T Y$ result in new variables given as linear combinations of $X$ and $Y$, without loss of generality, let us consider bivariate outcomes $X_i = (X_{1i}, X_{2i})^T$ and $Y_j = (Y_{1j}, Y_{2j})^T$, and set $l_1 = l_2 = (1,0)^T$ and $l_3 = l_4 = (0,1)^T$ in (19). Delong et al. [16] extended Sen's variance estimate (18) to covariance estimates for correlated variables. Specifically, following Delong's approach, equations in (17) can be redefined as

$$S_{10}^{kl} = \frac{1}{n_1 - 1} \sum_{i=1}^{n_1} \left(V_{10}(X_{ki}) - \hat{\varsigma}_{X_k,Y_k}\right)\left(V_{10}(X_{li}) - \hat{\varsigma}_{X_l,Y_l}\right),$$
$$\text{and } S_{01}^{kl} = \frac{1}{n_2 - 1} \sum_{j=1}^{n_2} \left(V_{01}(Y_{kj}) - \hat{\varsigma}_{X_k,Y_k}\right)\left(V_{01}(Y_{lj}) - \hat{\varsigma}_{X_l,Y_l}\right) \text{ for } k,l = 1,2, \quad (21)$$

where $V_{10}(X_{ki}) = \frac{1}{n_2}\sum_{j=1}^{n_2} \phi(X_{ki}, Y_{kj})$ for $i = 1,...,n_1$, $V_{01}(Y_{kj}) = \frac{1}{n_1}\sum_{i=1}^{n_1} \phi(X_{ki}, Y_{kj})$ for $j = 1,...,n_2$, and $\hat{\varsigma}_{X_k,Y_k}$ as in (15). Then the variance estimate of $\hat{\delta}$ can be given as

$$\hat{V}(\hat{\delta}) = (S_{10}^{11} - 2S_{10}^{12} + S_{10}^{22})/n_1 + (S_{01}^{11} - 2S_{01}^{12} + S_{01}^{22})/n_2. \quad (22)$$

Based on Theorem 1 and (22), we have the following result:

*Corollary 2: Under* $H_0 : \delta = \delta_0$, $-2\log R(\delta_0) \dfrac{\sum_{i=1}^{n_1}\sum_{j=1}^{n_2}(\phi_{12_{ij}} - \phi_{34_{ij}} - \delta_0)^2}{(n_1 n_2)^2 \hat{V}(\hat{\delta})}$ *converges in distribution to* $\chi_1^2$ *distribution.*

3.3 Comparison of two survival curves

A few generalizations of the two-sample Wilcoxon test statistic given censored data are available [17]. The most commonly used generalization is Gehan's test [18], which is known to have a higher power than the log rank test when the proportional hazards assumption is violated [19]. Suppose that we observe $(X_1, C_1), \cdots, (X_{n_1}, C_{n_1})$ that are $n_1$ independent observations from population 1 and $(X_{n_1+1}, C_{n_1+1}) \cdots, (X_{n_1+n_2}, C_{n_1+n_2})$ that are $n_2$ independent observations from population 2, where $C_i$ is 1 if the observations are censored, 0 if the observations are uncensored. We assume non-informative censoring. Gehan's test statistic estimates the difference between the two distributions (say $\tau$), which can be expressed in the form of $\hat{\tau} = \sum_{i=1}^{n_1} U_i$ where

$$U_i = \sum_{j=1}^{n_1+n_2} \phi_{ij},$$

$$\phi_{ij} = I\{(X_i, 0) > (X_j, 0) \text{ or } (X_i, 1) \geq (X_j, 0)\} - I\{(X_i, 0) < (X_j, 0) \text{ or } (X_i, 1) \leq (X_j, 0)\},$$

and the inequality is applied only to the first element. Let Gehan's test be expressed as

$$\hat{\tau} = \sum_{i=1}^{n_1} \sum_{j=n_1+1}^{n_1+n_2} \phi_{ij}. \tag{23}$$

Under $H_0: \tau = \tau_0 = 0$, the variance of $\hat{\tau}$ is estimated by

$$\hat{V}(\hat{\tau}) = \frac{n_1 n_2}{(n_1+n_2)(n_1+n_2-1)} \sum_{i=1}^{n_1+n_2} U_i^2.$$

We can construct the EL ratio test $R(\tau_0)$, similarly to (16) and (20), where now $w_{ij}$'s satisfy

$$Sup\left\{\prod_{i=1}^{n_1}\prod_{j=1}^{n_2} w_{ij} : \sum_{i=1}^{n_1}\sum_{j=n_1+1}^{n_2} w_{ij} = 1, \sum_{i=1}^{n_1}\sum_{j=n_1+1}^{n_2} w_{ij}\phi_{ij} = 0\right\}.$$

Since the statistic (23) is not divided by the number of summands, we have the following result:

*Corollary 3:* Under $H_0 : \tau = 0$, $-2\log R(\tau_0) \dfrac{\sum_{i=1}^{n_1}\sum_{j=n_1+1}^{n_2} \phi_{ij}^2}{\hat{V}(\hat{\tau})}$ converges in distribution to $\chi_1^2$ distribution.

### 3.4 Multivariate rank-based tests

An extension of the univariate rank procedures to multivariate problems can be carried out in a straightforward manner using Theorem 2. Particular definitions of the kernel function $h(\mathbf{X}, \mathbf{Y})$ can give rise to various U-statistic-based multivariate tests (e.g., see [20]). In this subsection, we illustrate an application of the proposed method for multivariate variables via a simple extension of the univariate Wilcoxon-Mann-Whitney statistic to that for multivariate variables. Consider independent continuous random vectors $X_i = (X_{1i}, \ldots, X_{pi})^T$, $i = 1, \ldots, n_1$ from population 1 and $Y_j = (Y_{1j}, \ldots, Y_{pj})^T$, $j = 1, \ldots, n_2$ from population 2. A null hypothesis in consideration is

$$H_0 : (p(X_1 < Y_1), \ldots, p(X_p < Y_p))^T = (0.5, \ldots, 0.5)^T. \qquad (24)$$

The probabilities in the left-hand side in the equation (24) are estimated by $\hat{\varsigma} = (\hat{\varsigma}_{X_1,Y_1}, \ldots, \hat{\varsigma}_{X_p,Y_p})$ using the definition (15). By the definitions in (21), let

$$\frac{1}{n_1}\begin{pmatrix} s_{10}^{kk} & s_{10}^{kl} \\ s_{10}^{lk} & s_{10}^{ll} \end{pmatrix} + \frac{1}{n_2}\begin{pmatrix} s_{01}^{kk} & s_{01}^{kl} \\ s_{01}^{lk} & s_{01}^{ll} \end{pmatrix} = \begin{pmatrix} C_{kk} & C_{kl} \\ C_{lk} & C_{ll} \end{pmatrix}. \qquad (25)$$

The covariance matrix of $\hat{\varsigma}$ is estimated by

$$\hat{V}(\hat{\varsigma}) = (C_{ij}), i, j = 1, \ldots, p,$$

where $(C_{ij})$ is a matrix with $C_{ij}$ in (25) for its $(i,j)$-th element. Let $\varsigma_0 = (0.5,...,0.5)^T$. Under $H_0$ in (24), we can construct the EL ratio test $R(\varsigma_0)$ similarly to (16). Specifically, $w_{ij}$'s satisfy

$$Sup\left\{\prod_{i=1}^{n_1}\prod_{j=1}^{n_2} w_{ij} : \sum_{i=1}^{n_1}\sum_{j=n_1+1}^{n_2} w_{ij} = 1, \sum_{i=1}^{n_1}\sum_{j=n_1+1}^{n_2} w_{ij}(\phi_{ij} - \varsigma_0) = 0\right\},$$

where $\phi_{ij} = (\phi(X_{i1}, Y_{j1}),..., \phi(X_{ip}, Y_{jp}))^T$ by the definition used in (15) and $0$ is the $p \times 1$ zero vector. Let

$$\hat{H} = \frac{1}{n_1 n_2}\sum_{i=1}^{n_1}\sum_{j=1}^{n_2}(\phi_{ij} - \varsigma_0)(\phi_{ij} - \varsigma_0)^T,$$

which is a nonsingular matrix. Let $\lim_{n\to\infty, n_1/n \to r} \hat{H}^{-1}\hat{V}(\hat{\varsigma}) = K$. Then, by Theorem 2, we have the following result:

*Corollary 4: Under $H_0$ in (24), $-2\log R(\varsigma_0)/(n_1 n_2)$ converges in distribution to $\sum_{k=1}^{p} c_k \chi_{1k}^2$ where $c_k$'s are the eigenvalues of $K$.*

The relevant R codes for the methods proposed in this section are available at a website (http://sphhp.buffalo.edu/biostatistics/faculty-and-staff/faculty-directory/jinheeyu.html). Simulation results of all the tests developed in this section follow in the next section.

## 4. Simulation Study

We carry out a Monte Carlo study (5,000 repetitions per scenario) for the tests proposed in Section 3 under various combinations of underlying distributions, which reflect a violation of the exchangeability assumption [7] under $H_0$. The significance level is fixed at 0.05 throughout the simulations.

For the ROC curve simulation study, we compare the proposed test statistic in Corollary 1 to the test statistics proposed by the plug-in method [2], jackknife method [1], and the normal approximation based on Sen's variance estimator [15]. Our simulation results show that the EL approach is generally more robust (in terms of the Type I error) to various underlying distributions and different values of the AUC than other approaches. A few simulation results are presented in Figure 1, where the assumption of exchangeability is not met for the underlying distributions. The results presented are based on distributions of $N(0,1)$ vs. $N(\mu, 2^2)$ (Normal vs. Normal), Lognormal$(0,1)$ vs. Lognormal$(\mu, 2^2)$ (Lognormal vs. Lognormal), and Lognormal$(0,1)$ vs. Normal$(\mu, 2^2)$ (Lognormal vs. Normal) where the relevant $\mu$ is selected to obtain the AUC's of 0.8, 0.9 and 0.95 under $H_0$. As shown in Figure 1, our approach has more robust Type I error control than the existing methods (i.e., the Type I error is close to the nominal Type I error), especially under various underlying distributions with higher AUC values (e.g, 0.95). Across the other methods, Type I error control seems unstable in some common settings (e.g., normal distributions with AUC=0.95). This fact implies that inferences based on those methods may be unreliable. We note that the jackknife approach may be too conservative when the sizes of the two groups are different, and the convergence of the test may be too slow with respect to increasing sample sizes. Through the Monte Carlo study, with unequal sample sizes, we observe that the data generated by the jackknife generally violate the basic assumption of the classical one-sample EL approach (e.g., observations are from an identical distribution), which may have given rise to its poor performance. With unequal sample sizes, the asymptotic result of the jackknife approach (i.e., Theorem 2 in [1]) may require substantially larger sample sizes in terms of convergence than the sample sizes presented in this article.

An examination of the power of the various tests under the same distributional families used to generate Figure 1 is presented in Figure 2. The power properties are examined through increasing values of $\mu$. The shift values for $\mu$ are 1 for Normal vs. Normal, 1 for Lognormal vs. Lognormal and 2 for Lognormal vs. Normal. When the sample size is small, some of the tests have a greater power than the proposed method. However, this is most likely due to their uncontrolled Type I error rates as seen in Figure 1. In some large sample sizes cases, we note that the normal approximation may have an advantage in terms of the power over the proposed approach, where the Type I error rates are reasonably controlled. However, as shown in the case of lognormal vs. normal with sample sizes of $n_1 = n_2 = 100$, the normal approximation does not have satisfactory Type I error control compared to the proposed approach, even with relatively large sample sizes. Overall, the proposed approach is more robust toward maintaining the specified Type I error rates in decision making under various circumstances. This fact is particularly important in the nonparametric setting where various underlying distributions are possible in real data analysis and are generally unknown.

We also carry out a simulation study that compares two correlated ROC curves (relative to Corollary 2) based on various underlying distributions and AUC values. The proposed method is compared to the approach based on a normal approximation [16]. Overall, both the normal approximation and the proposed method show reasonable Type I error control in many of the considered cases; however, the proposed method maintains the nominal Type I errors across different values of AUC better than the other approach. A few cases showing this property are presented in Figure 3. The power comparison for these tests is also presented in Figure 4. In the figures, paired observations used for these simulations are distributed as $\begin{pmatrix} X_1 \\ X_2 \end{pmatrix} \sim N_2 \left( \begin{pmatrix} 0 \\ 0 \end{pmatrix}, \begin{pmatrix} 1 & 0.9 \\ 0.9 & 1 \end{pmatrix} \right)$ vs.

$$\begin{pmatrix}Y_1\\Y_2\end{pmatrix}\sim N_2\left(\begin{pmatrix}\mu_1\\\mu_2\end{pmatrix},\begin{pmatrix}4&3.6\\3.6&4\end{pmatrix}\right),\begin{pmatrix}X_1\\X_2\end{pmatrix}\sim N_2\left(\begin{pmatrix}0\\0\end{pmatrix},\begin{pmatrix}1&1.8\\1.8&4\end{pmatrix}\right)\text{ vs. }\begin{pmatrix}Y_1\\Y_2\end{pmatrix}\sim N_2\left(\begin{pmatrix}\mu_1\\\mu_2\end{pmatrix},\begin{pmatrix}4&7.2\\7.2&16\end{pmatrix}\right),\begin{pmatrix}\log X_1\\\log X_2\end{pmatrix}\sim N_2\left(\begin{pmatrix}0\\0\end{pmatrix},\begin{pmatrix}1&0.9\\0.9&1\end{pmatrix}\right)\text{ vs.}$$

$$\begin{pmatrix}\log Y_1\\\log Y_2\end{pmatrix}\sim N_2\left(\begin{pmatrix}\mu_1\\\mu_2\end{pmatrix},\begin{pmatrix}4&3.6\\3.6&4\end{pmatrix}\right),\text{ and }\begin{pmatrix}\log X_1\\\log X_2\end{pmatrix}\sim N_2\left(\begin{pmatrix}0\\0\end{pmatrix},\begin{pmatrix}1&1.8\\1.8&4\end{pmatrix}\right)\text{ vs. }\begin{pmatrix}\log Y_1\\\log Y_2\end{pmatrix}\sim N_2\left(\begin{pmatrix}\mu_1\\\mu_2\end{pmatrix},\begin{pmatrix}4&7.2\\7.2&16\end{pmatrix}\right),\text{ indicating the}$$

cases of the 'same' normal distributions, 'different' normal distributions, the 'same' lognormal distributions and 'different' lognormal distributions, respectively. For the Type I error examination (Figure 3), the values of $\mu_i$ are selected to achieve the true AUC's of 0.8, 0.9 and 0.95. For the simulated power (Figure 4), the location parameters used for the figure are $(\mu_1+0.4,\mu_2)^T$ for the same normal, different normal, the same lognormal and different lognormal distributions, respectively. As demonstrated in Figure 3, the proposed method controls Type I error rates well with the small sample sizes and higher AUC values (e.g., 0.95), while both methods are comparable for other cases. In general, the power of the proposed method responds well when sample sizes are increased, and is comparable to the normal approximation method as shown in Figure 4.

The robust characteristic of our new approach in terms of Type I error control is shown with the two-group survival analysis test (relative to Corollary 3) as well. For our simulations, we consider cases with non-constant hazard functions based on different *Weibull* distribution parameterizations. Under the alternative hypothesis, we consider cases that violate the proportional hazards assumption. Various censoring rates are used based on the Poisson arrival of individuals and differing lengths of follow-up times. Some examples' results are presented in Figure 5. The results in Figure 5 are based on a comparison of two survival curves with the distributions of *Weibull*(1,1) vs. *Weibull*(1,1) for the Type I error, and *Weibull*(0.5,1) vs. *Weibull*(2,1) and *Weibull*(0.5,1) vs. *Weibull*(5,1) for the power. Also, each individual is assumed to enter the study based on the standard exponential distribution (parameter = 1), and overall censoring rate set at 20%. Overall, the performance of the proposed method is similar to Gehan's test; that is, both Gehan's test and the proposed method show comparable Type I error rate control and comparable values for power.

Finally, we investigate the Type I error and the power of the multivariate extension of the Wilcoxon-Mann-Whitney test (relative to Corollary 4). For our simulations, we compare two bivariate random variables where the probabilities $P(X_1 < Y_1)$ and $P(X_2 < Y_2)$ based on the marginal distributions strictly maintain 0.5 under $H_0$. Some results are presented in Figure 6. For the multinormal distributions in Figure 6, we present $\begin{pmatrix} X_1 \\ X_2 \end{pmatrix} \sim N_2 \left( \begin{pmatrix} 0 \\ 0 \end{pmatrix}, \begin{pmatrix} 4 & 1.5 \\ 1.5 & 2.25 \end{pmatrix} \right)$ vs. $\begin{pmatrix} Y_1 \\ Y_2 \end{pmatrix} \sim N_2 \left( \begin{pmatrix} 0 \\ 0 \end{pmatrix}, \begin{pmatrix} 25 & 2.5 \\ 2.5 & 1 \end{pmatrix} \right)$ for investigating the Type I error rates. The means of $Y_1$ and $Y_2$ are replaced by 0.5, respectively, for the simulated power. For the multivariate lognormal distributions in Figure 6, we present $\begin{pmatrix} \log X_1 \\ \log X_2 \end{pmatrix} \sim N_2 \left( \begin{pmatrix} 0 \\ 0 \end{pmatrix}, \begin{pmatrix} 4 & 1.5 \\ 1.5 & 2.25 \end{pmatrix} \right)$ vs. $\begin{pmatrix} \log Y_1 \\ \log Y_2 \end{pmatrix} \sim N_2 \left( \begin{pmatrix} 0 \\ 0 \end{pmatrix}, \begin{pmatrix} 1 & 0.25 \\ 0.25 & 0.25 \end{pmatrix} \right)$ for the Type I error, and the means of $Y_1$ and $Y_2$ are replaced by 0.5 for the power. For the comparison, a test statistic based on the normal theory is used, i.e., $(\hat{\varsigma} - \varsigma_0)^T \hat{V}(\hat{\varsigma})^{-1} (\hat{\varsigma} - \varsigma_0)$ which has a $\chi^2$ distribution with the two degrees of freedom under $H_0$ (the chi-squared test in Figure 6). As shown in Figure 6, the proposed method has a better Type I error control across the different sample sizes. With small sample sizes, the power of the $\chi^2$ test in Figure 6 may look like an improvement as compared to the proposed method; however, it should be more beneficial to use the proposed method since the $\chi^2$ test has somewhat inflated Type I errors in the small sample size setting.

**5. An Application to Crossover Designs**

In this section, we demonstrate the diverse applicability of the proposed methods by applying the proposed approach to data analyses in the context of crossover designs. Crossover designs are commonly used in areas with outcomes that evaluate responses with relatively short term effects and chronic conditions. To name a few, we observe that crossover designs are used to study pharmacokinetic parameters (e.g., [21]), performance of cognitive tasks (e.g., finger tapping in [22]) and air quality improvement [23]. We note that the crossover designs may have a methodological limitation in handling the long-term carryover effect. More detailed discussions are found in [24] and [25].

Although crossover designs are commonly used in bioequivalence studies based on pharmacokinetic parameters, the results derived under normality assumptions can be affected by some outlying observations, which in turn may lead to incorrect decisions regarding the bioequivalence of two or more agents [26]. Due to the robustness of our approach to outliers and parametric assumptions, we propose it as a strong alternative to normal based tests in this setting. Herein, we apply the proposed methods to the 2×2 crossover design and the 2×2 crossover design with baseline and washout period data. The washout period is defined to be the time between treatment periods to remove the carryover effects from the previous treatment [27].

In this article, an application of the proposed methods is illustrated using the peak heart rate data analyzed by Jung and Koch [27]. The peak heart rate data are based on a study comparing a novel treatment (say B) and an active control (say A) for patients with ventricular arrhythmia and organic heart disease [27]. In the study, a total of 20 patients were randomly assigned to the two sequence groups (9 in the A:B group and 11 in the B:A group). The response variable in the data set was the peak heart rate during the bicycle exercise test in four different periods (baseline, first treatment, washout and second treatment). We note that the sample size may not be sufficiently large; however, we use this data set for the purpose of demonstrating that the proposed methods can be easily adapted for diverse sets of hypothesis tests.

The 2×2 crossover design [28] compares two treatments in the form of two different sequences (A:B and B:A) during two periods. Let $Y_{ijk}$ be the outcome variable where subscript $i$ is 1 for the sequence A:B and 2 for the sequence B:A, subscript $j$ is for the period (1 or 2), and subscript $k$ indicates independent subjects. Using the similar notations of Kenward and Jones [29], the response in the 2×2 crossover design is expressed as:

$$Y_{11k} = \mu - \gamma + \pi_1 - \tau + \varepsilon_{11k}, Y_{21k} = \mu + \gamma + \pi_1 + \tau + \varepsilon_{21k},$$
$$Y_{12k} = \mu - \gamma + \pi_2 + \tau - \theta + \varepsilon_{12k}, Y_{22k} = \mu + \gamma + \pi_2 - \tau + \theta + \varepsilon_{22k}, \qquad (26)$$

where $\mu, \gamma, \pi_j, \tau$ and $\theta$ are the overall mean, the sequence effect, the $j$-th period effect (where $\pi_1 + \pi_2 = 0$), the treatment effect and the carryover effect, respectively, and $\varepsilon_{ijk}$ is an independent random effect with a mean of 0 and the same distribution within the sequence $i$ and period $j$. For notational ease, let $Y_{ijk} \sim Y_{ij}$. In case of an equal carryover effect (including a case of no carryover effect) between the two sequences, $\theta$ is 0 since $\pi_2$ can replace the carryover effect in such cases. The standard procedure of hypothesis testing consists of two steps:

Step 1. Test the carryover effect.

Step 2. If the carryover effect is significant at the first step, the treatment effect is tested based only on the outcomes of the first period. Otherwise, the treatment effect is tested based on the outcomes of all periods.

The 'relative effect' between two groups can be defined by a probability $P(Z_i > Z_j)$, where $Z_i$ indicates the outcome variable from Group $i$ [30]. No difference between the two treatment groups is expressed by $P(Z_i > Z_j) = 0.5$, which is the null hypothesis of the Wilcoxon-Mann-Whitney test [31, 32]. The probability $P(Z_i > Z_j)$ is estimated as a summation of relevant ranks [31]. In the crossover design, the treatment effects are tested based on contrasts of $Y_{ij}$'s (e.g., [27]). Note that general linear combinations for means may not be translated to linear combinations of the probabilities or ranks [20]; however, the probabilities to compare different $Y_{ij}$'s can be used to produce a test statistic relevant to a null hypothesis of interest when the probability of the relative treatment effect can be clearly specified under the null hypothesis.

For the first step, the carryover effect is commonly tested based on the null hypothesis $H_0 : E(Y_{11} + Y_{12}) = E(Y_{21} + Y_{22})$ or $H_0 : E(Y_{11} - Y_{21}) = E(Y_{22} - Y_{12})$, which can be interpreted that the relative effect between $Y_{11}$ and $Y_{21}$ is same as that between $Y_{22}$ and $Y_{12}$. In this spirit, a null hypothesis can be rewritten as

$$H_0 : P(Y_{11} > Y_{21}) = P(Y_{22} > Y_{12}), \tag{27}$$

based on the definition of the relative effect. The hypothesis (27) can be tested by the direct application of Corollary 2, which tests two correlated probabilities. For the second step, on the decision of no carryover effect, we look into the treatment effect in both periods. The null hypothesis to indicate no relative treatment effect in either of the two periods is

$$H_0 : P(Y_{11} > Y_{21}) = 0.5 \text{ and } P(Y_{12} > Y_{22}) = 0.5. \tag{28}$$

On the other hand, upon deciding the existence of the carryover effect, the treatment effect is tested based only on the first period as

$$H_0 : P(Y_{11} > Y_{21}) = 0.5. \tag{29}$$

The both hypotheses (28) and (29) are tested by the direct application of Corollary 4.

Based on the peak heart rate data set from the first and second treatment periods, the estimated values of $P(Y_{11} > Y_{21})$ and $P(Y_{22} > Y_{12})$ are 0.732 and 0.899, respectively. We note that, since there are many tied values to calculate $I(Y_{ijk} > Y_{i'j'k'})$ in the peak heart rate data set, the value of 0.5 is assigned in cases of tied values, and the variances are estimated accordingly. The test statistic to test the carryover effect has the value of 1.0889 (p-value =0.298 based on the $\chi_1^2$ distribution), indicating no strong evidence of the carryover effect. Thus, we test the treatment effect based on the hypothesis (28) as the next step. The estimated values of $P(Y_{11} > Y_{21})$ and $P(Y_{12} > Y_{22})$ are 0.727 and 0.101, respectively, showing sizable deviations from the probability 0.5 in the both periods. The log-likelihood ratio statistic $-2\log R(\varsigma_0)$ in Corollary 4 is 929.538, and the corresponding Monte-Carlo p-value based on the estimated variances (i.e., $\hat{H}$ and $\hat{V}(\hat{\varsigma})$) is smaller than 0.0001, strongly indicating a treatment difference. We note for readers that there are some discussions related to keeping the size of the overall test under the nominal value [25,33] for this multiple testing scheme in the crossover designs, although it may not be of specific interest of this article since our analysis example is more towards illustrating the testing methodology.

Now let us consider the 2×2 crossover design with baseline and washout periods. Herein, we follow the analytical scheme of Kenward and Jones [29] to illustrate how hypothesis tests can be carried out using the proposed methods. We note that some discussions indicate that the usage of differences between baseline and treatment outcomes may decrease the study power in some cases [25]. Nevertheless, it is also common that the change from the baseline is of specific interest in many studies (e.g., [34]). In such cases, it may be more desirable to use the differences themselves as outcome variables. Due to the baseline and washout periods, besides the responses in (26), we have additional observations as

$$X_{11k} = \mu - \gamma + \pi_b + \varepsilon_{1bk}, \quad X_{21k} = \mu + \gamma + \pi_b + \varepsilon_{2bk},$$
$$X_{12k} = \mu - \gamma + \pi_w - \lambda + \varepsilon_{1wk}, \quad X_{22k} = \mu + \gamma + \pi_w + \lambda + \varepsilon_{2wk},$$

where the subscripts $b$ and $w$ indicate the effects in baseline and washout periods, $\lambda$ indicates the first order carryover effect, and $\sum_{i \in \{1,2,b,w\}} \pi_i = 0$. Note that Jung and Koch [27] considered $\gamma = 0$ in their analysis. The value $\theta$ in the responses (26) now indicates the second order carryover effect. Three steps of tests are now the part of our testing procedure: 1) the first order carryover effect, 2) the second order carryover effect and 3) the treatment effect. The ordinary least squares estimator of the first order carryover effect has the structure of the contrast of the observations, $\{X_{11} - X_{21} - (X_{12} - X_{22})\}/2$, which allows us to test the null hypothesis $H_0: \lambda = 0$ [29]. This contrast can be interpreted as comparing the relative effect between $X_{11}$ and $X_{21}$ and that between $X_{12}$ and $X_{22}$. Thus, the null hypothesis can be rewritten as

$$H_0: P(X_{11} > X_{21}) = P(X_{12} > X_{22}), \tag{30}$$

which can be tested by the direct application of Corollary 2. When we fail to reject the hypothesis (30), the second order carry-over effect is tested as the next step. Let us define $Z_{ij} = Y_{ij} - X_{ij}$, the difference between the treatment and baseline (or washout) for the respective sequence group $i$ and period $j$. The second order carryover effect is tested based on the contrast of the observations, $\{Z_{11} - Z_{21} - (Z_{22} - Z_{12})\}/2$, which is indeed the estimator of the treatment-by-period interaction [29]. The corresponding null hypothesis can be rewritten as

$$H_0: P(Z_{11} > Z_{21}) = P(Z_{22} > Z_{12}), \tag{31}$$

which is equivalent to $H_0 : \theta = 0$ ($\theta$ as the second carryover effect), and can be tested by an application of Corollary 2. Upon deciding that there is no carryover effect, the treatment effect is suggested to be tested based on the hypothesis (28). If the either (30) or (31) is rejected, the test is based on the first period only as

$$H_0 : P(Z_{11} > Z_{21}) = 0.5.$$

For the heart peak data set, the estimated values of $P(X_{11} > X_{21})$ and $P(X_{12} > X_{22})$ are exactly the same 0.293, indicating no first order carryover effect. The corresponding test statistic is 0 since the maximum of the EL function under $H_0$ is same as that under $H_1$. For the second order carryover effect, the estimated values of $P(Z_{11} > Z_{21})$ and $P(Z_{22} > Z_{12})$ are 0.919 and 0.778, respectively. The test statistic is 1.196 (p-value = 0.274) indicating no strong evidence of the second order carryover effect. Upon deciding no first or second order carryover effect, the treatment effect is tested based on the hypothesis (28), which we already presented above.

## 6. Concluding Remarks

In this article, it was shown that U-statistics can be used as constraints in the framework of the EL approach in a two-group comparison setting. The limiting distribution of the EL likelihood ratio test has a weighted $\chi^2$ distribution or a combination of weighted $\chi^2$ distributions. We demonstrated that the application of the proposed approach to any U-statistic type of tests can be carried out in a straightforward manner. With relatively small sizes, we observed a fairly good approximation of the proposed tests through robust Type I error control. The proposed approaches are generally more robust than the other test statistics based on U-statistics for various situations, including a violation of exchangeability under the null hypotheses.

One reason for such robustness is that the proposed approaches are in essence based on the principle of the ratio of the empirical likelihood function, which accommodate different shapes of underlying distributions better than the normal approximation-based U-statistics that rely on symmetry of the limiting distribution. Another reason for the accuracy of the proposed method in the approximation, particularly as compared to other EL approaches can be due to a quicker convergence rate of the Lagrange multiplier $\lambda^*$ in (12). In a typical one-sample EL problem or two-sample EL problem, the Lagrange multiplier convergence rate is the inverse of the square root of the overall sample size or the sample size of one group (e.g., [1,2]); however, in our approach for two-groups, the convergence rate is based on the product of two group sample sizes giving rise to a fast convergence (for more details, see the Appendix). Our approach is conceptually simpler than other EL approaches, as those approaches require creating new summands for the constraints either by using pseudo-values or incorporating other sample statistics into the constraints (plug-in approach). The simplicity of our approach comes with some expense of a larger number of weight estimates (i.e., $w_{ij}$'s); however, estimating weights is simply a matter of obtaining the solution of $\lambda^*$ in the nonlinear equation (11) and using the relationship (10), which hardly requires extra effort. We also presume that the simplicity of our approach may contribute to some improvement of our approach as compared to the other EL approaches.

Overall, the results presented in this article demonstrate that an EL approach based on U-statistic constraints has workable properties and may improve the performance of traditional U-statistics approximation.

## Appendix

We need the following lemma for the proof of Theorem 1.

**Lemma 1** [13, 35]

(i) Assume that $E|h(\mathbf{X},\mathbf{Y})| < \infty$. Then $\hat{\theta} \xrightarrow{a.s.} \theta$ as $n \to \infty$.

(ii) Assume that $Eh^2(\mathbf{X},\mathbf{Y}) < \infty$. Suppose $\frac{n_1}{n_1+n_2} \to r$ as $n \to \infty$, where $n = n_1 + n_2$, then we have $\sqrt{n}(\hat{\theta} - \theta) \xrightarrow{d} N(0, v^2)$, where $v^2 = \frac{m_1^2}{r}\sigma_{1,0}^2 + \frac{m_2^2}{1-r}\sigma_{0,1}^2$ as in (5).

*Proof of Theorem 1.*

Since $\min_{i,j \in S} h(\mathbf{X}_i, \mathbf{Y}_j) < \theta < \max_{i,j \in S} h(\mathbf{X}_i, \mathbf{Y}_j)$, there is a solution of $\lambda^*$ that satisfies

$$\frac{1}{N}\sum_{i,j \in S}\frac{h(\mathbf{X}_i,\mathbf{Y}_j) - \theta}{1 + \lambda^*(h(\mathbf{X}_i,\mathbf{Y}_j) - \theta)} = 0, \tag{A.1}$$

where $N = \binom{n_1}{m_1}\binom{n_2}{m_2}$. Let $\psi_{ij}^c = h(\mathbf{X}_i, \mathbf{Y}_j) - \theta$. From (A.1), we have

$$0 = \frac{1}{N}\left|\sum_{i,j \in S}\psi_{ij}^c - \lambda^*\sum_{i,j \in S}\frac{(\psi_{ij}^c)^2}{1 + \lambda^*\psi_{ij}^c}\right| \geq \frac{|\lambda^*|}{N}\sum_{i,j \in S}\frac{(\psi_{ij}^c)^2}{1 + \lambda^*\psi_{ij}^c} - \left|\sum_{i,j \in S}\frac{\psi_{ij}^c}{N}\right|$$

$$\geq \frac{|\lambda^*|}{1 + |\lambda^*|K}\left[\sum_{i,j \in S}\frac{(\psi_{ij}^c)^2}{N}\right] - \left|\sum_{i,j \in S}\frac{\psi_{ij}^c}{N}\right|, \tag{A.2}$$

where $K = \max_{i,j}\{|\psi_{ij}^c| : i,j \in S\}$. It can be shown that $K = o(N^{1/\alpha})$ since $E(h(\mathbf{X}_i, \mathbf{Y}_j))^\alpha < \infty$ for some integer $\alpha (\geq 2)$ [10,36].

By Lemma 1, the second term of the right-side of the inequality (A.2) is $O_p(n_1^{-1/2})$, and the value inside the bracket in the first term is $O(1)$, that is, it has the upper bound less than $\infty$ as it is a U-statistic and $E(h(\mathbf{X}_i, \mathbf{Y}_j) - \theta)^2 < \infty$. Thus, $|\lambda^*|/(1+|\lambda^*|K) = O_p(n_1^{-1/2})$, and it follows that

$$|\lambda^*| = O_p(N^{-1/2}). \tag{A.3}$$

Expanding (A.1), we have

$$0 = \frac{1}{N}\sum_{i,j\in S}\psi_{ij}^c - \frac{\lambda^*}{N}\sum_{i,j\in S}(\psi_{ij}^c)^2 + \frac{(\lambda^*)^2}{N}\sum_{i,j\in S}\frac{(\psi_{ij}^c)^3}{1+\lambda^*\psi_{ij}^c}. \tag{A.4}$$

Note $\max_{i,j}\{\lambda^*|\psi_{ij}^c|\} = O_p(N^{-1/2})o(N^{1/\alpha}) = o_p(1)$ for $\alpha \geq 2$. And,

$$\sum_{i,j\in S}\frac{(\psi_{ij}^c)^3}{N} \leq K\sum_{i,j\in S}\frac{(\psi_{ij}^c)^2}{N} = o(N^{1/\alpha})O(1) = o(N^{1/\alpha}). \tag{A.5}$$

Thus, the last term of (A.4) is $O_p(N^{-1})o(N^{1/\alpha})o_p(1) = o_p(N^{-1/2})$ for $\alpha \geq 2$. This leads to

$$\lambda^* = \sum_{i,j\in S}\psi_{ij}^c / \sum_{i,j\in S}(\psi_{ij}^c)^2 + o_p(N^{-1/2}). \tag{A.6}$$

Now,

$$l(\theta_0) = -2\log R(\theta_0) = -2\left[\sum_{i,j\in S}(\log w_{ij} + \log N)\right] = 2\left[\sum_{i,j\in S}\log\{1+\lambda^*\psi_{ij}^c\}\right]$$

$$= 2\left[\sum_{i,j\in S}\lambda^*\psi_{ij}^c - \frac{1}{2}\sum_{i,j\in S}\lambda^{*2}(\psi_{ij}^c)^2 + \frac{1}{3}\sum_{i,j\in S}|\lambda^*|^3 \xi(\psi_{ij}^c)^3\right], \tag{A.7}$$

for some $\xi < \infty$, thus the last term is $O_p(N^{-3/2})o(N^{1+1/\alpha}) = o_p(1)$ by letting $\alpha = 2$ and using (A.3) and (A.5). By (A.6) and (A.7),

$$-2\log R(\theta_0) = 2\cdot\frac{\left[\sum_{i,j\in S}\psi_{ij}^c\right]^2}{\sum_{i,j\in S}(\psi_{ij}^c)^2} - \frac{\left[\sum_{i,j\in S}\psi_{ij}^c\right]^2}{\sum_{i,j\in S}(\psi_{ij}^c)^2} + o_p(1) = \frac{(\hat{\theta}-\theta)^2\cdot N^2}{\sum_{i,j\in S}(\psi_{ij}^c)^2} + o_p(1).$$

Since $\sum_{i,j \in S} (\psi_{ij}^c)^2 / N$ is a consistent estimator of $\eta^2$, by Lemma 1 and Slutsky's theorem,

$$-2 \log R(\theta_0) \cdot \frac{\eta^2}{N \cdot \sigma^2} \xrightarrow{d} \chi_1^2.$$

We remark that the first term of the right-hand side of the equation (A.6) multiplied by $\binom{n_1}{m_1}\binom{n_2}{m_2}$ is used for the initial value to obtain the solution of (11).

*Proof of Theorem 2.*

The proof of Theorem 2 is similar to that of Theorem 1, thus we only give an outline below. Let $K = \max_{i,j} \{|d^T \psi_{ij}^c| : i, j \in S\}$ for a fixed vector $d$. Since $E|d^T h(\mathbf{X}, \mathbf{Y})|^\alpha < \infty$, $K = o(N^{1/\alpha})$ for some integer $\alpha (\geq 2)$ [10,36]. Similarly to (A.2), we expand $\frac{1}{N} \left| \sum_{i,j \in S} \frac{d^T \psi_{ij}^c}{1 + \lambda^* \psi_{ij}^c} \right|$ which leads to $\|\lambda^*\| = O_p(N^{-1/2})$. Also, expanding the left-hand side of the equation (14) similarly to (A.4) and solving for $\lambda^*$, we have $\lambda^* = \left( \sum_{i,j \in S} \psi_{ij}^c \psi_{ij}^{c\,T} \right)^{-1} \sum_{i,j \in S} \psi_{ij}^c + o_p(N^{-1/2})$. Expanding $l(\theta_0)$ in (13) similarly to (A.7), plugging $\lambda^*$ and using $\|\lambda^*\| = O_p(N^{-1/2})$, $l(\theta_0)$ asymptotically has the form of

$$N \left( \sum_{i,j \in S} \psi_{ij}^c / N \right)^T \Sigma^{-1/2} \Sigma^{1/2} \left( \sum_{i,j \in S} \psi_{ij}^c \psi_{ij}^{c\,T} / N \right)^{-1} \Sigma^{1/2} \Sigma^{-1/2} \left( \sum_{i,j \in S} \psi_{ij}^c / N \right),$$

where $\sum_{i,j \in S} \psi_{ij}^c \psi_{ij}^{c\,T} / N$ is the unbiased U-statistic of $H$, and $\Sigma^{-1/2} \left( \sum_{i,j \in S} \psi_{ij}^c / N \right)$ has $q$-variate multinormal distribution $N_q(0, I)$. Thus, the desired result follows.

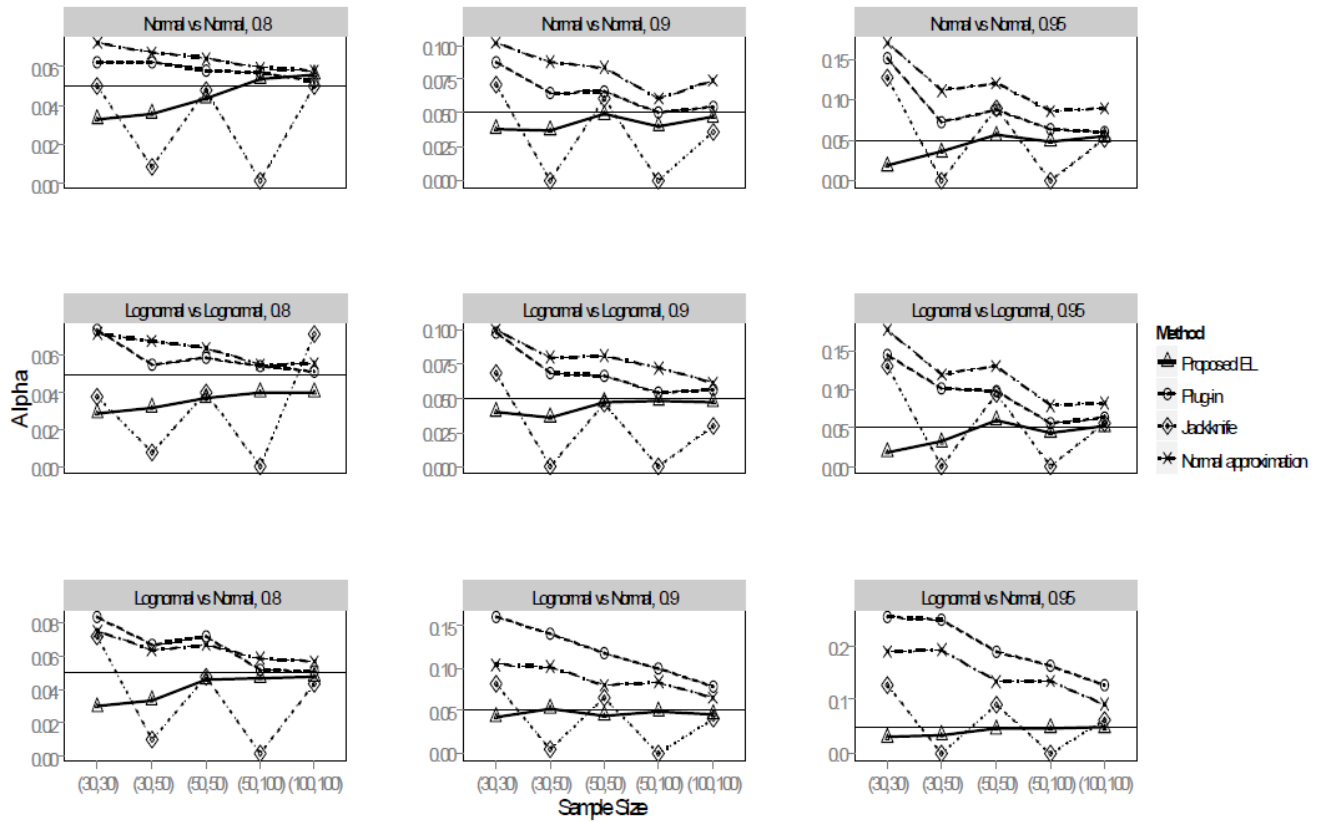

Figure 1: Simulated Type I errors of the ROC analysis with various underlying distributions. The columns from the left in the figure represent the AUC of 0.8, 0.9 and 0.95, respectively, and each row represents different combinations of the distributions as stated at the top of each plot. Parameters used are described in Section 4. The sample sizes are shown on the bottom plots. The horizontal solid line in each plot indicates 0.05, the target Type I error.

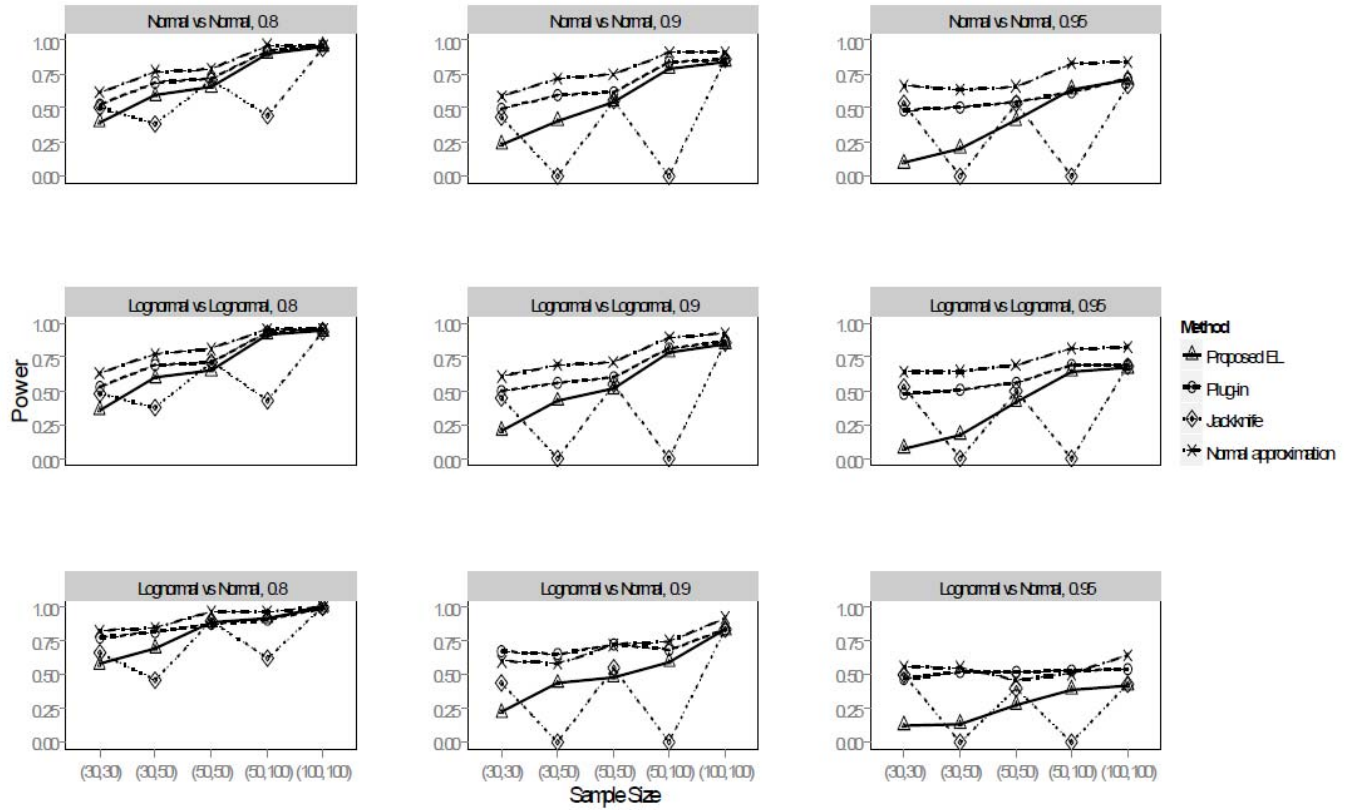

Figure 2: Simulated powers of the ROC analysis with various underlying distributions. The columns from the left in the figure represent the AUC of 0.8, 0.9 and 0.95, respectively, and each row represents different combinations of the distributions as stated at the top of each plot. Parameters used are described in Section 4. The sample sizes are shown on the bottom plots.

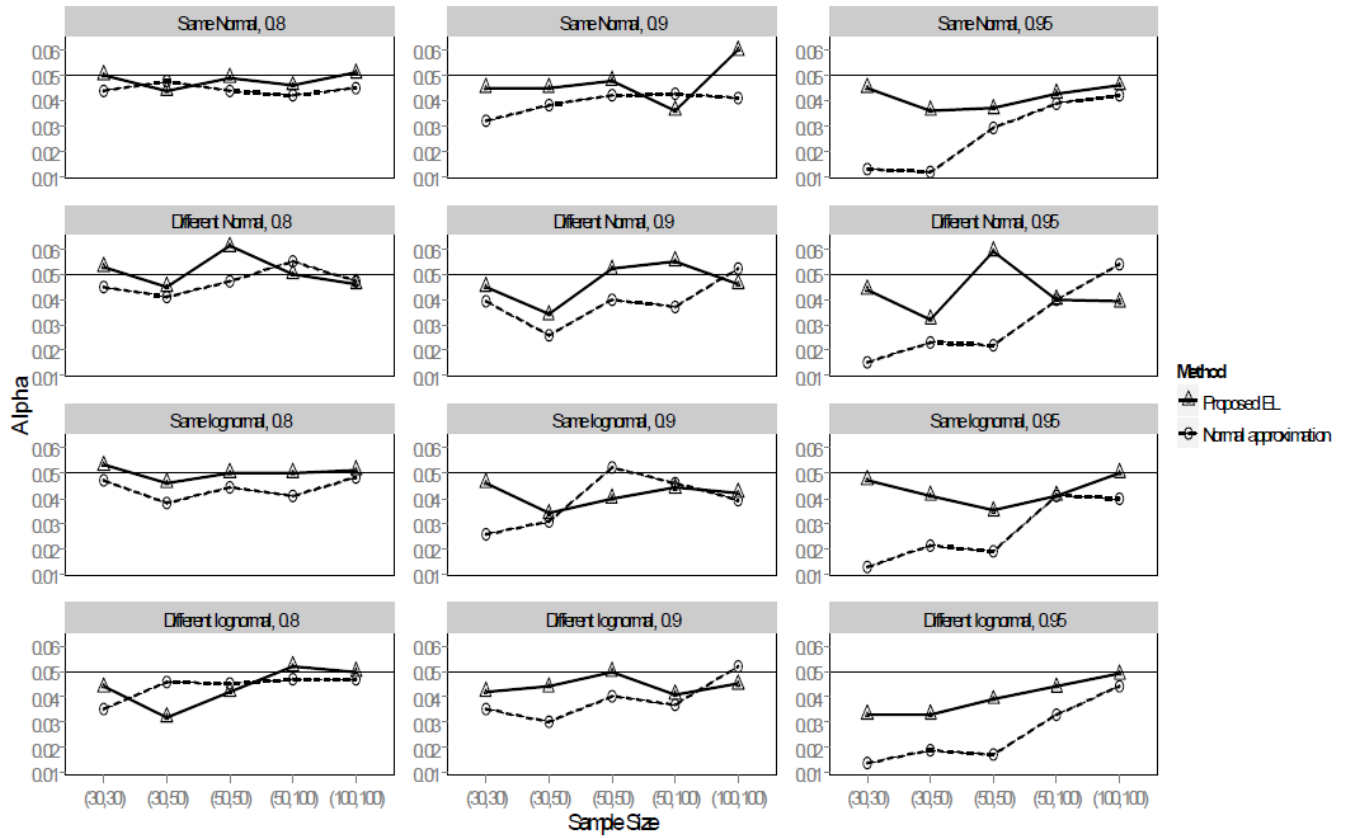

Figure 3: Simulated Type I errors of the comparison of ROC curves with the various underlying distributions. The columns from the left in the figure represent the AUC of 0.8, 0.9 and 0.95, respectively, and each row represents different combinations of the distributions as stated at the top of each plot. Parameters and distributions used are described in Section 4. The horizontal solid line in each plot indicates 0.05, the target Type I error. The sample sizes are shown on the bottom plots.

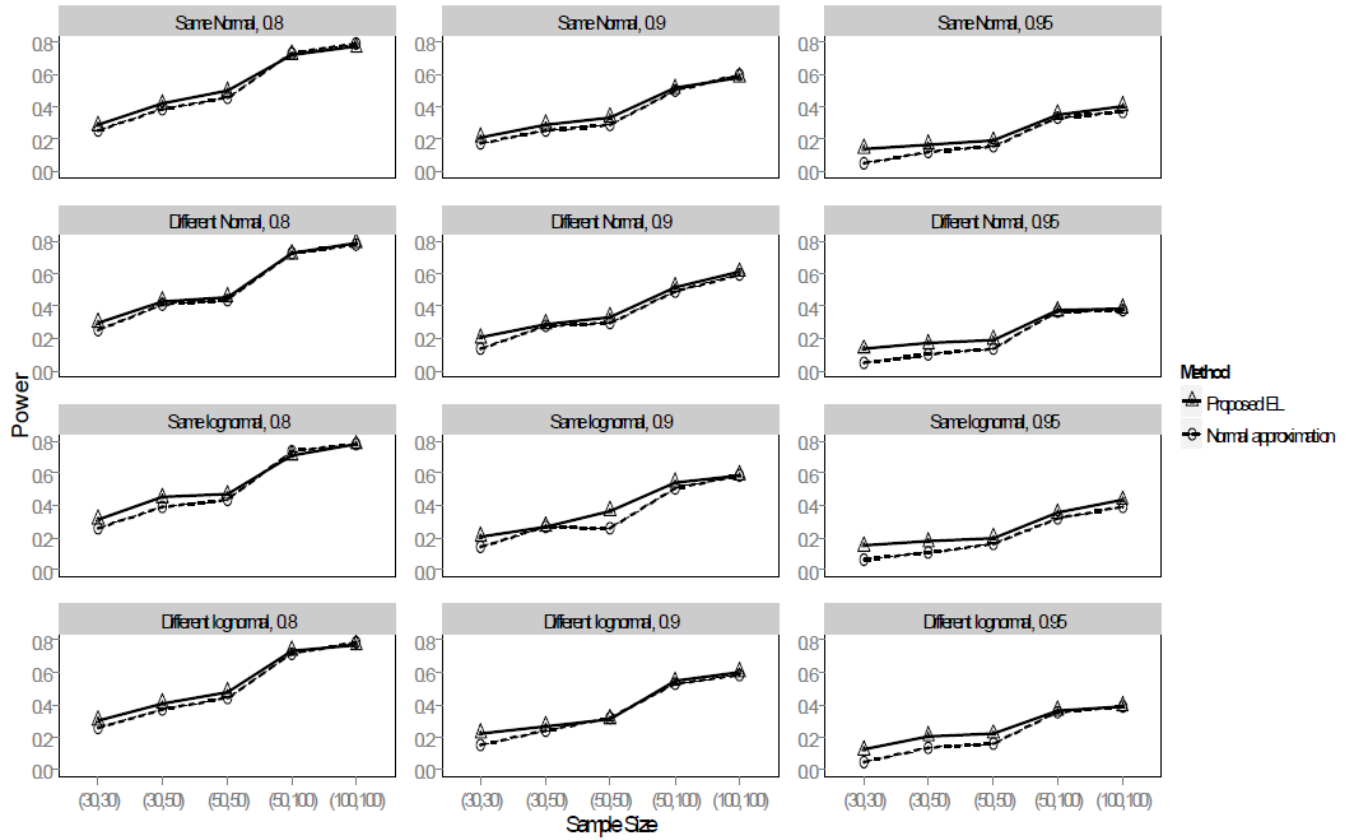

Figure 4: Simulated powers of the comparison of ROC curves with the various underlying distributions. The columns from the left in the figure represent the AUC of 0.8, 0.9 and 0.95, respectively, and each row represents different combinations of the distributions as stated at the top of each plot. Parameters and distributions used are described in Section 4. The sample sizes are shown on the bottom plots.

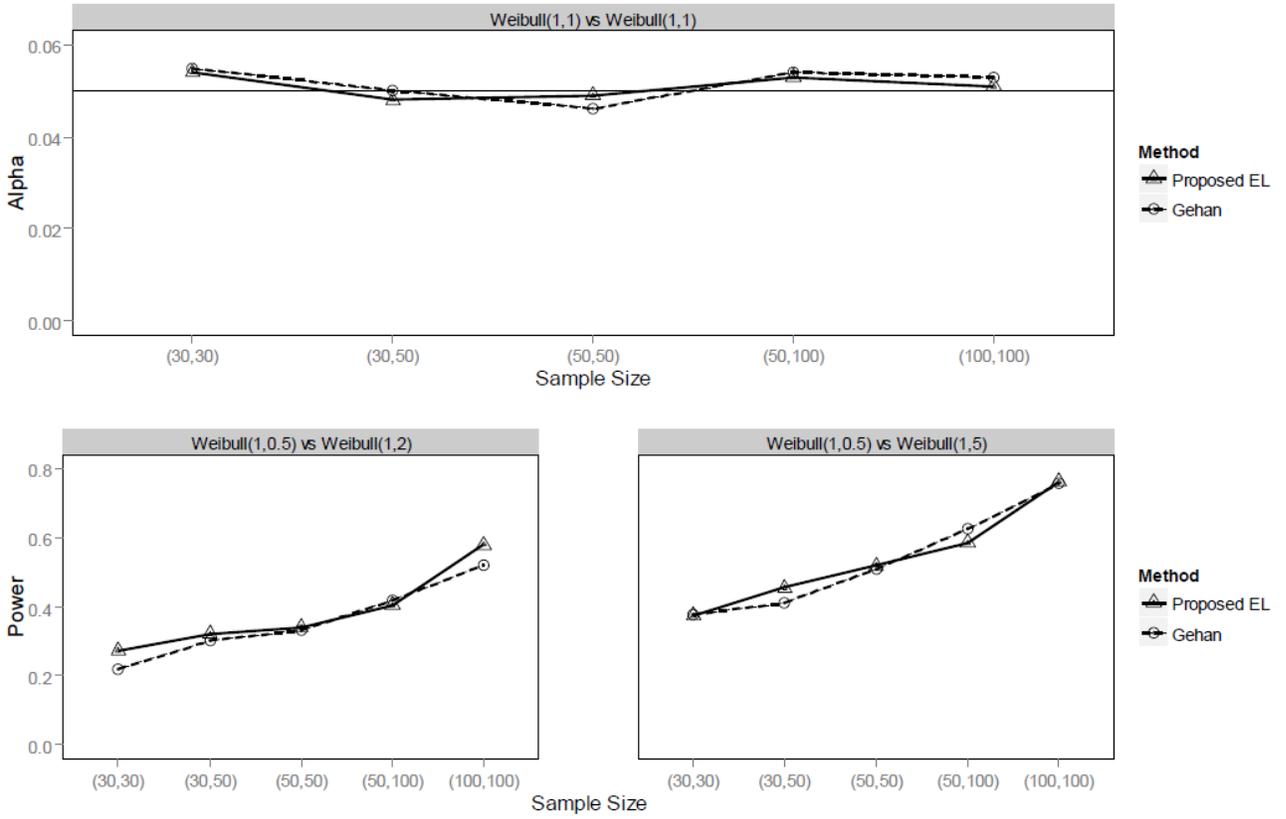

Figure 5: Simulated Type I errors (first row) and powers (second row) for two-group survival comparisons. The target Type I error is 0.05 (solid horizontal line in the plot on the top).

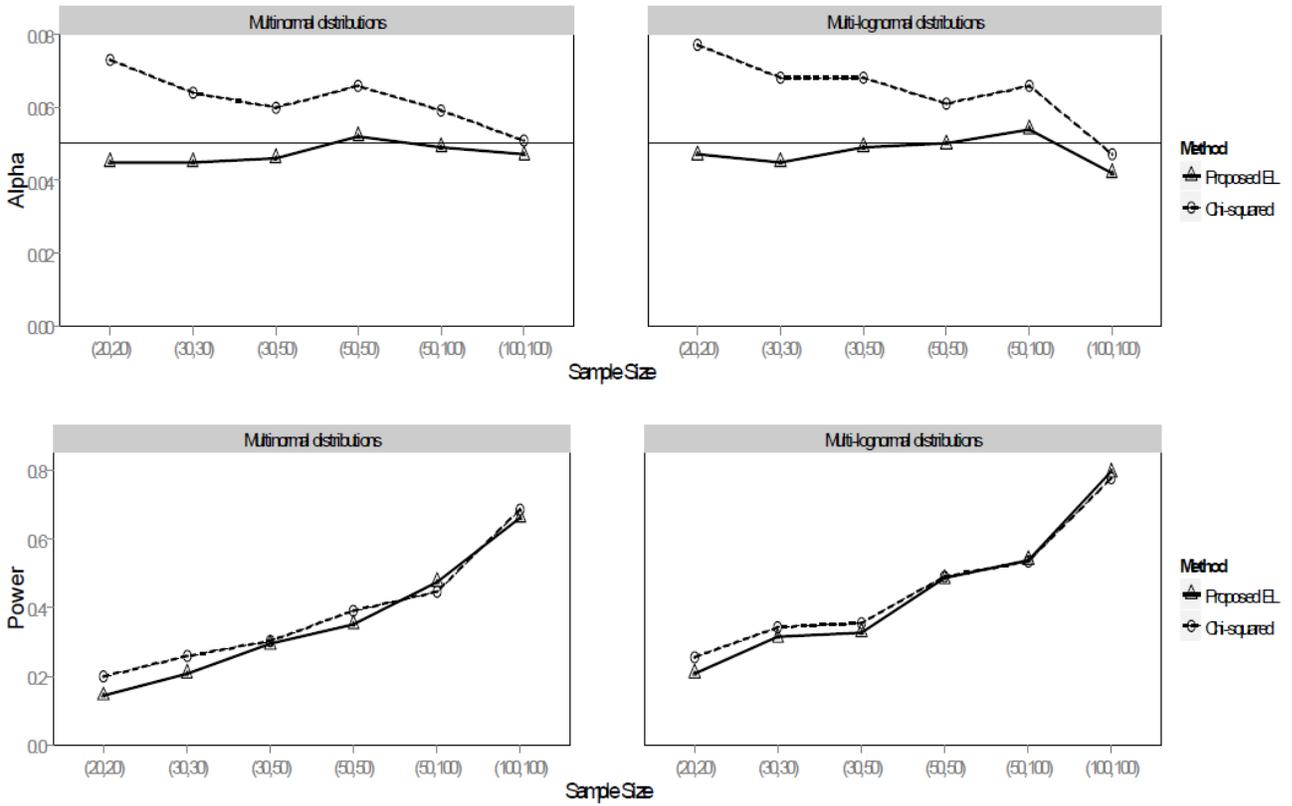

Figure 6: Simulated Type I errors (first row) and powers (second row) for the multivariate Wilcoxon-Mann-Whitney test. The target Type I error is 0.05 (solid horizontal line in the plots at the first row).